\documentclass[prl,twocolumn]{revtex4-1}

\usepackage{amsmath}    
\usepackage{graphicx}
\usepackage{verbatim}   
\usepackage{color}      
\usepackage{subfigure}  
\usepackage{hyperref}   

\begin{document}

\title{First-order phase transition and tricritical point in multiband $U(1)$ London superconductors}

\author{Karl A. H.  Sellin$^{1}$ and Egor Babaev$^{1}$}
\affiliation{$^1$Department of Theoretical Physics  and  Center for Quantum Materials, The Royal Institute of
  Technology, SE-10691 Stockholm, Sweden}  
  
\date{\today}
\begin{abstract}
The order of the superconducting phase transition 
 is a classical  problem.
Single-component  type-2 superconductors
 exhibit a continuous ``inverted-XY" phase transition,
as was first demonstrated
for $U(1)$ lattice London superconductors by a celebrated duality mapping with
subsequent backing by numerical simulations.
Here we study this problem in multiband $U(1)$ London
superconductors and find evidence that by contrast the model has a tricritical point. The
superconducting
phase transition becomes first-order when the Josephson length is sufficiently large
compared to the
magnetic field penetration length. We present evidence  that the fluctuation-induced
dipolar interaction between vortex loops makes
the phase transition discontinuous. 
We discuss that this mechanism
is also relevant for the  phase transitions
  in  multicomponent gauge theories  with higher broken symmetry.

\end{abstract}
\maketitle

The problem of the order of the superconducting phase transition beyond mean-field approximation is more than 40 years old.
In \cite{PhysRevLett.32.292} it was observed that if phase fluctuations are neglected in the Ginzburg-Landau model, fluctuations of the vector potential make the superconducting phase transition first-order (see also \cite{PhysRevD.7.1888}).
This scenario applies only for strongly type-1 superconductors, as was demonstrated by later works \cite{Thomas1978513,Peskin1978122,PhysRevLett.47.1556}. These works
  considered a lattice London model, neglecting fluctuations of the modulus of the order parameter but taking into account phase fluctuations. It was demonstrated that the superconducting phase transition beyond mean-field approximation is caused by the proliferation of vortex loops with short-range interaction set by the magnetic field penetration length $\lambda$. For low temperatures
only small vortex rings are thermally excited, while at $T_c$ the thermally excited vortex loops loose their line tension and extend throughout the entire system.

In Refs. \cite{Thomas1978513,Peskin1978122,PhysRevLett.47.1556} 
 a duality mapping was established between the statistical problem of the normal state above $T_c$ in a superconductor
and a statistical description of a superfluid state below $T_c$, such that the temperature axis is reversed, see also \cite{PhysRevLett.80.1964}. Thus the phase transition in the lattice London superconductor is called the
``inverted-3D XY" transition. The London limit approximation is justifiable for extreme type-2 superconductors. { The question at what parameter values of the Ginzburg-Landau model the inverted-3D XY phase transition turns into first-order has subsequently been studied using various approaches.
Some of the attempted analytical approaches   \cite{kleinert1982disorder}  suggested that the continuous phase transition extends slightly into the region
where the Ginzburg-Landau parameter $\kappa$ is smaller than $1$ (or, equivalently, in traditional units smaller than $1/\sqrt{2}$). However this approach
was based on uncontrollable assumptions, and thus numerical simulations were required to address that question.  Early numerical work were consistent with the existence of a tricritical point \cite{bartholomew1983phase}, as well as one-loop renormalization-group calculations \cite{PhysRevLett.76.4588}. The largest-scale Monte Carlo simulation performed so far \cite{PhysRevB.65.104501} suggests that the continuous phase transition indeed extends slightly into the region of $\kappa<1$. All these works suggested that in London model (which corresponds to taking
extremely type-2 limit) the phase transition is continuous.}

Recently there has been a surge of interest to multicomponent  $U(1)\times U(1)$ 
and $SU(2)$ generalizations of this problem in the context of multicomponent 
superconducting condensates  \cite{babaev2004phase,babaev2004superconductor,PhysRevB.82.134511}
and of deconfined quantum criticality proposals where such Ginzburg-Landau models
were argued to arise as an effective field theory  
\cite{PhysRevB.70.075104,Senthil05032004,sachdev2008quantum,chen2013deconfined}. Initially it was suggested that $SU(2)$ as well as $U(1)\times U(1)$
superconductors with equal phase stiffnesses possess a continuous phase transition
from a state with fully broken symmetries to a normal state in a novel universality class  \cite{Senthil05032004,PhysRevB.70.075104,motrunich2008comparative},
but later analysis revealed first-order phase transitions  \cite{cond-mat/0501052,PhysRevLett.101.050405,chen2013deconfined,Kuklov20061602,PhysRevB.82.134511,
0805.2578,2006PhRvL..97x7201K,
herland2013phase} in such systems. The flowgram method proposed in \cite{Kuklov20061602}
shows that the phase transition remains first order in the limit of infinitesimally small electric charge. 
First-order phase transitions were also found
in $U(1)\times Z_2$ systems, where a $Z_2$ symmetry associated with 
broken time-reversal symmetry is caused by a frustrated intercomponent coupling \cite{PhysRevB.89.104509}.
 
Most of the superconductors which are of great current interest are multiband superconductors
 \cite{PhysRevLett.3.552,moskalenko1959superconductivity,
leggett1966number}. In the London
limit multiband superconductors are described by several phases coupled by a
Josephson interaction \cite{leggett1966number}. Here we consider the problem of the phase transition in a  multi-component London superconductor with condensates $\psi_{a}=\vert\psi_a\vert\exp(i\theta\sp{(a)})$ ($a=1,2,...$), with constant $\vert\psi_a\vert$, described by the free-energy density
\begin{align}
	f=&\frac{1}{2}\sum_{a}
	\vert\psi_a\vert^2
	(\nabla\theta^{(a)}+e\mathbf{A})\sp2
	+
	\frac{1}{2}
	\left(\nabla\times\mathbf{A}\right)^2\nonumber\\
	&-
	\sum_{a>b}
	\eta_{ab}	
	\vert\psi_a\vert\vert\psi_b\vert
	\cos(\theta^{(a)}-\theta^{(b)}),
	\label{model}
\end{align}
where $e$ is the electric charge coupling and $\eta_{ab}$ determines the strength of the Josephson phase-difference-locking term between bands $a$ and $b$. We are interested here in the case where
the Josephson coupling breaks the symmetry explicitly to $U(1)$, and to restrict the parameter space we will consider $\eta_{ab}=\eta$.
{ In \cite{smiseth2005field} a duality mapping was discussed for a class of $U(1)$ multi-band models. Namely, it was discussed that the dual version of 
the model is described by proliferation of a single kind of directed loops. That indicates that if the phase transition is continuous then it should be
in the ``inverted-3D XY'' universality class. However it was established recently that intermediate-length-scale interaction
in the directed-loops model can make the phase transition first-order  \cite{meier2015fluctuation}. Thus the  origin of the phase transition in multi-band models requires a careful study.}

We begin by examining the two-band case, then
without loss of generality $\eta$ can be taken to be non-negative, such that in the ground state
$\theta^{(1)}-\theta^{(2)}=0$.
 From the free energy two characteristic length scales can be identified:
 the first one is the London magnetic penetration depth $\lambda$ and the second one is the Josephson length $\xi_J$, the latter at which the system restores the ground state from small deviations in the phase difference.
 As previously mentioned, the London model is justified for strongly type-2 multiband superconductors, where the coherence lengths associated with the densities are much smaller than $\lambda$  and $\xi_J$. The Josephson length $\xi_J$ can be identified by writing
$\theta\sp{(1)}(x)-\theta\sp{(2)}(x)=\delta(x)$, imposing the condition
$\delta(0)=\delta_0$, and expanding the Josephson term. 
Then the   system recovers ground state value of the phase difference according to the exponential law
$
\delta(x)=\delta_0\exp\left(-x/\xi_J\right)
$
with
$
\xi_J=
\sqrt{
\vert\psi_1\vert\vert\psi_2\vert
/
(\eta
(\vert\psi_1\vert\sp2+\vert\psi_2\vert\sp2))
}
$. The
standard expression for the magnetic field penetration depth is
$\lambda
=1/(e\sqrt{\vert\psi_1\vert\sp2+\vert\psi_2\vert\sp2})
$.

When $\eta=0$, the model has $U(1)\times U(1)$ symmetry. 
As mentioned above, the phase diagram of such a system have been previously studied, in two dimensions \cite{babaev2004phase}, three dimensions
in an external field \cite{babaev2004superconductor,
smorgrav2005observation,
smorgrav2005vortex} and three-dimensional cases without an external field \cite{Kuklov20061602,PhysRevB.82.134511}. In three dimensions, at a low but non-zero value of the coupling constant $e$, the model exhibits
a single first-order phase transition from the $U(1)\times U(1)$-state to the normal state.
For large electric charges there are two phase transitions. The reason for occurrence of the second
phase transition is the following:
at a lower critical temperature a proliferation of bound states of vortices with winding
in both phases takes place: i.e. vortices for which  $\Delta\theta\sp{(1)}\equiv\oint_\sigma \nabla \theta \sp{(1)}=2\pi,\ 
\Delta\theta\sp{(2)}\equiv\oint_\sigma \nabla \theta\sp{(2)}=2\pi$ 
 where the integration path $\sigma$
encloses two cores, such vortices are denoted by (1,1). The ``elementary" vortices
in the individual condensates are bound into these composite objects through the coupling
to the vector potential. In the resulting state the individual phases $\theta\sp{(a)}$
are disordered but the phase difference is ordered \cite{babaev2004phase,babaev2004superconductor,smorgrav2005observation,Kuklov20061602,PhysRevB.82.134511}. This state is called a metallic superfluid, paired state or super-counter-fluid.  
At elevated temperatures the transition into the normal state is driven by the proliferation of 
 individual (also termed fractional) vortices  $\Delta\theta\sp{(1)}\equiv\oint_\sigma \nabla \theta\sp{(1)}=2\pi,\ 
\Delta\theta\sp{(2)}\equiv\oint_\sigma \nabla\theta \sp{(2)}=0$, denoted by $(1,0)$ and  
$\Delta\theta\sp{(1)}\equiv\oint_\sigma \nabla \theta\sp{(1)}=0,\ 
\Delta\theta\sp{(2)}\equiv\oint_\sigma \nabla \theta\sp{(2)}=2\pi$, denoted by $(0,1)$. For the properties of the individual and 
composite vortices in this model see  \cite{PhysRevLett.89.067001,babaev2004phase}.

It has been discussed that the existence of the paired
 states in multicomponent gauge theories could be related to first-order phase transitions  \cite{Kuklov20061602}.
 Indeed, even at the level of mean-field analysis the direct $U(1)\times U(1)$ or $SU(2)$ phase transitions are first-order in the vicinity (on the phase diagram) of the paired state. However mean-field analysis
 is qualitatively wrong for this problem in general, failing to capture first order phase transition happening
 in the $e\to 0+$ limit.
 By contrast, the model (\ref{model}) shares the same symmetry as  single-component superconductors, indeed the Josephson coupling breaks the symmetry explicitly to $U(1)$.
  
By discretizing the model  \eqref{model} onto a cubic grid of size $L$, rewriting the gradients with finite differences, the kinetic energy terms into XY-model cosines, and inserting the previously identified length scales, we obtain the lattice Hamiltonian for the  two-band case
\begin{align}
	H=\sum_{i}\Bigg[	
&\sum_{a}\sum_{\mu}
-\vert\psi_a\vert^2
\cos(\theta^{(a)}_{i+\hat{\mu}}
-\theta^{(a)}_i+A_{i\mu})\nonumber\\
	&+
	\frac{1}{2}	\lambda^2(\vert\psi_1\vert\sp2+\vert\psi_2\vert\sp2)\sum_{\mu}
	[\Delta\times\mathbf{A}]_{i\mu}^2	
	+\nonumber\\
	&-
	\frac{1}
	{\xi_J^2}	
	\frac{\vert\psi_1\vert\sp2\vert\psi_2\vert\sp2}
	{\vert\psi_1\vert^2+\vert\psi_2\vert^2}
	\cos(\theta^{(1)}_i-\theta^{(2)}_i)\Bigg],
	\label{eq:Hamiltonian}
\end{align}
where $i$ is a lattice site index, $\mu=x,y,z$ and $[\Delta\times\mathbf{A}]_{i\mu}=A_{i\mu'}+
A_{i+\hat{\mu}',\mu''}
-A_{i+\hat{\mu}'',\mu'}
-A_{i\mu''}$ is the discrete lattice curl, where we have denoted $x'=y$, $y'=z$ and $z'=x$. We simulate the temperature-driven phase transitions of the Hamiltonian \eqref{eq:Hamiltonian} with a parallel tempering Metropolis Monte Carlo algorithm \cite{PhysRevLett.57.2607,B509983H,newman1999monte} and use periodic boundary conditions.
The trial moves consist of selecting a site at random and proposing new values for all five site degrees of freedom (two phases and three vector potential components). The phases are updated by drawing random numbers between 0 and $2\pi$, and the components of the vector potential are updated by adding (either positive or negative) random numbers 
to the old values. Parallel tempering swap trial moves between adjacent temperatures are attempted with a fixed frequency \cite{B509983H,1.477973} of once every sweep. To collect data we perform simulations of at least $10\sp6$ sweeps and we perform the simulations with linearly distributed temperatures. 

\begin{figure}
	\includegraphics[width=\columnwidth]{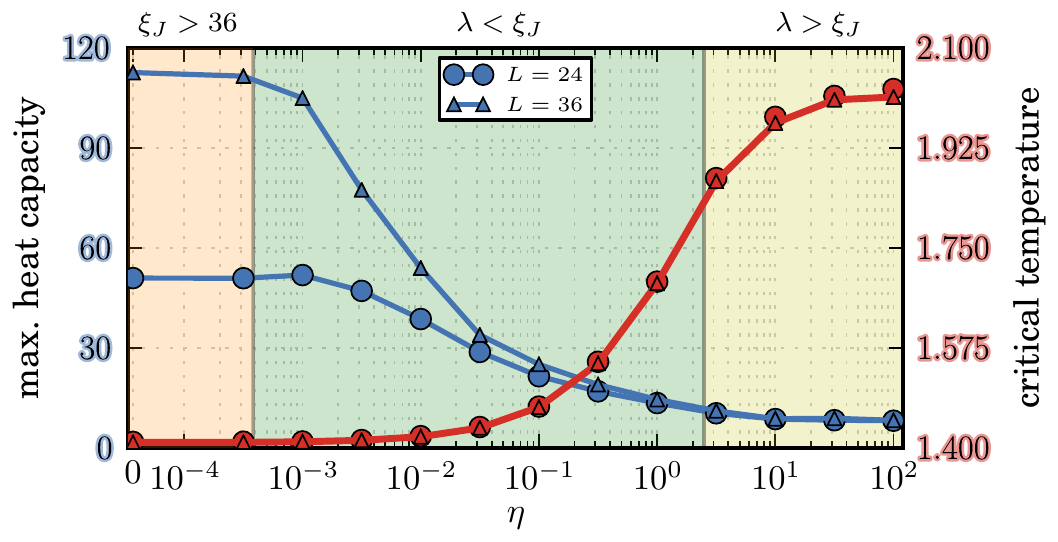}
	\caption{The maximum of the calculated heat capacity (left vertical axis, blue) and its corresponding critical temperature (right vertical axis, red) for $L=24$ (circles) and $L=36$ (triangles), with Josephson coupling $\eta$ varying over several orders of magnitude.}
	\label{fig:cmax_etavar}
\end{figure}

We begin by determining how a finite Josephson coupling term affects the maximal value of the heat capacity and the corresponding temperature. We consider the case of twin bands $\vert\psi_1\vert^2=\vert\psi_2\vert^2=1$ with  $\lambda\sp2=0.2$ 
and vary $\eta$ over several orders of magnitude. Fig. \ref{fig:cmax_etavar} shows the maximal value of the calculated heat capacity as well as the corresponding temperature, taken to be the critical temperature for two system sizes. Each point on the curves corresponds to a simulation and the lines are guides to the eye (note however that the temperature curves overlap). The simulations have a fixed set of temperatures each, chosen manually to cover the heat capacity peak.

As is seen in Fig. \ref{fig:cmax_etavar}, the heat capacity peak is larger in the region where the Josephson interaction is small. As the Josephson interaction is increased, the Josephson length becomes smaller as does the magnitude of the heat capacity peak. In the right-most region where the Josephson length is smaller than the magnetic penetration depth, the magnitude of heat capacity peak is not visibly different for the two system sizes. The critical temperature raises as the heat capacity peak diminishes and is not visibly affected by changing the system size. The behavior of the displayed quantities suggests that the left-most and right-most points of Fig. \ref{fig:cmax_etavar}) represents two distinct physical regimes.

For the left-most point in Fig. \ref{fig:cmax_etavar} (with $\eta=0$) the symmetry of the system is $U(1)\times U(1)$, which, as previously mentioned, is known to have a first-order phase transition \citep{cond-mat/0501052,Kuklov20061602,PhysRevB.77.144519}. This, combined with the observation that the heat capacity maximum grows substantially with the system size in the region of finite and small $\eta$, suggests that the phase transition is of first-order also for finite $\eta$ in the region with $\lambda<\xi_J$. In the region of $\lambda>\xi_J$ however, any growth in the heat capacity maximum with system size is not visible in Fig. \ref{fig:cmax_etavar}, suggesting there is a tricritical point and a continuous phase transition for $\lambda>\xi_J$.

\begin{figure}
	\includegraphics[width=\columnwidth]{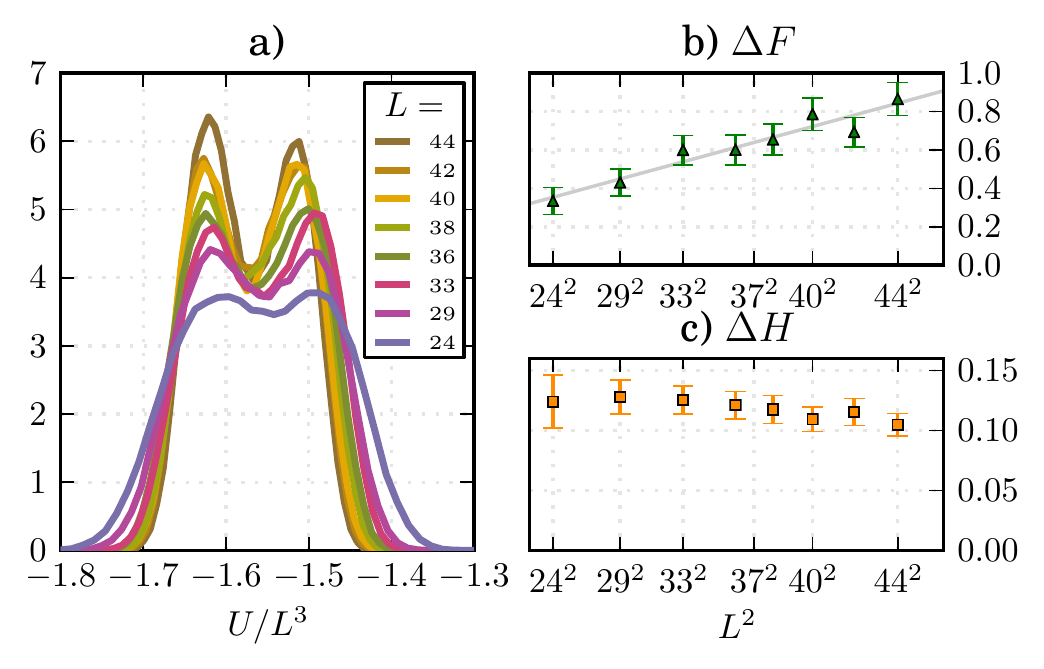}
	\caption{For $\eta=10\sp{-3}$ the scaling behavior of the energy histograms (normalized distributions of the internal energy per site $U/L^3$) at the phase transition is indicative of a first-order phase transition. The histograms for various lattice sizes are shown in a), the free-energy barrier $\Delta F$ is shown in b) and the latent heat $\Delta H$ in c).}
	\label{fig:histogramscaling}
\end{figure}

We consider first the scaling properties for $\eta=10\sp{-3}$ (with $\xi_J\approx23$) which is in the $\lambda<\xi_J$-region of Fig. \ref{fig:cmax_etavar}. The internal energy histogram at the phase transition is seen in Fig. \ref{fig:histogramscaling} a) to be bimodal, and the bimodality is enhanced by increasing the system size. The free-energy barrier $\Delta F=(\beta_c)\sp{-1} \ln(P_{max}/P_{min})$ ($\beta_c$ is the inverse critical temperature and $P$ the energy distribution) is seen in Fig. \ref{fig:histogramscaling} b) to be proportional to $L^2$ and the latent heat $\Delta H$ (i.e. the distance between the peaks) is seen in c) to not vanish, indicating a first-order phase transition \cite{PhysRevLett.65.137,PhysRevB.43.3265}. By contrast, we observed no bimodal features or tendencies for any system with $\lambda>\xi_J$.

\begin{figure}
	\includegraphics[width=\columnwidth]{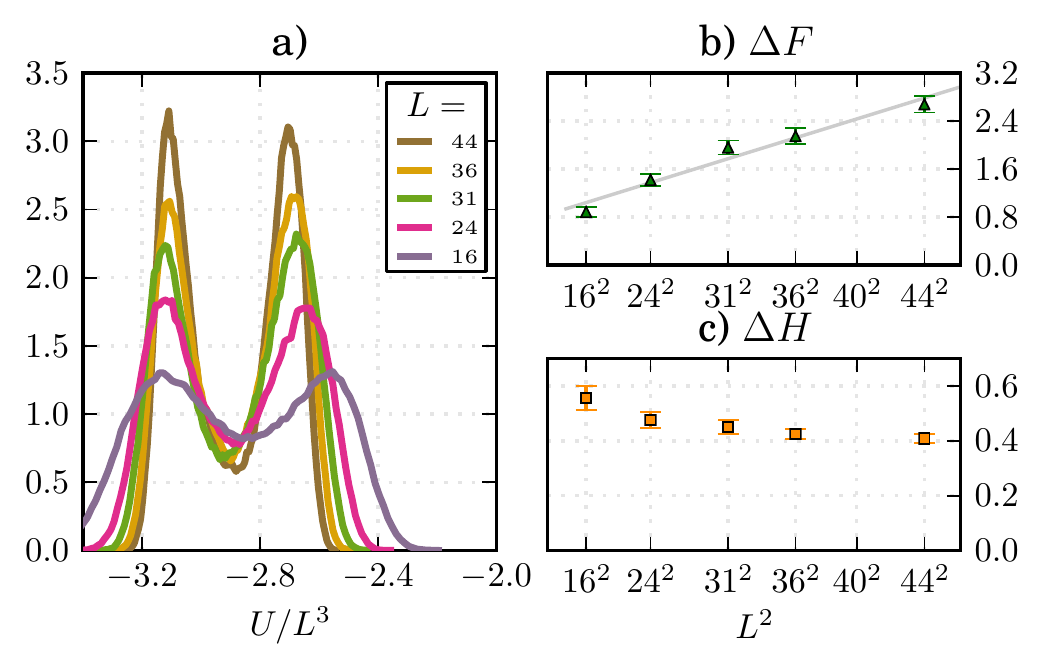}
	\caption{First-order scaling of the energy histograms for the three-band model also hold when $\xi_J\ll L$, here $\xi_J=3.3$ with scaling up to $L=44$.}
	\label{fig:histograms}
\end{figure}

\begin{figure}
	\includegraphics[width=\columnwidth]{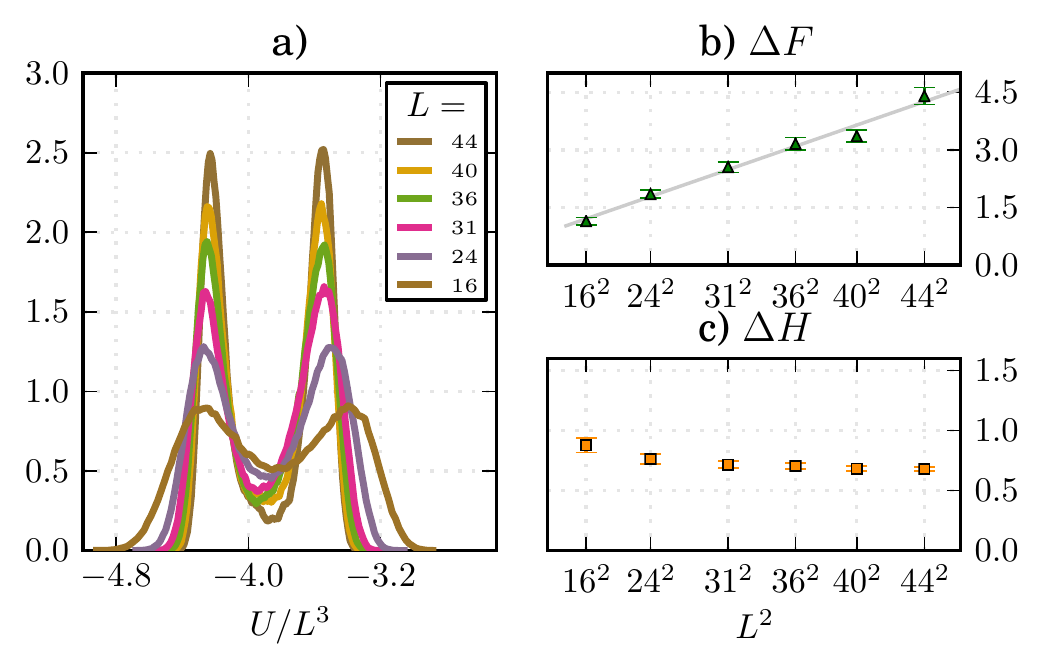}
	\caption{First-order scaling of the energy histograms is more pronounced for a four-band model.}
	\label{fig:histograms4band}
\end{figure}

 We find that when the Josephson interaction is increased, the system size required for a bimodal energy distribution to develop also increases. This makes it computationally difficult to reproduce the scaling results of Fig. \ref{fig:histogramscaling} for systems where the Josephson length is much smaller than the box length. To further test these phenomena we consider a three- and four-band models with $\vert\psi_i\vert\sp2=1$
with a Josephson coupling which locks all the phase differences to zero in the ground state.
In such a system we find that the first-order phase transition becomes stronger and the measurement of bimodal energy distributions for $\xi_J\ll L$ is possible. In Fig. \ref{fig:histograms} is shown the first-order scaling of the energy histograms for a three-band model with $\eta=0.03$ such that $\xi_J\approx 3.3$ up to $L=44$, more that ten Josephson lengths. The results of Fig. \ref{fig:histograms} are $10\sp6$ sweep simulations with $\lambda\sp2=0.1$. { In Fig. \ref{fig:histograms4band} is shown simulation results for a four-band model with an even stronger Josephson coupling of $\eta=0.07$, and $\lambda\sp2=0.075$. The four-band results are $2\cdot 10\sp6$ sweep simulations with sampling during the last $500 000$ sweeps. With the growing number of components we find stronger signatures of the first order phase transitions.}

\begin{figure}
	\includegraphics[width=\columnwidth]{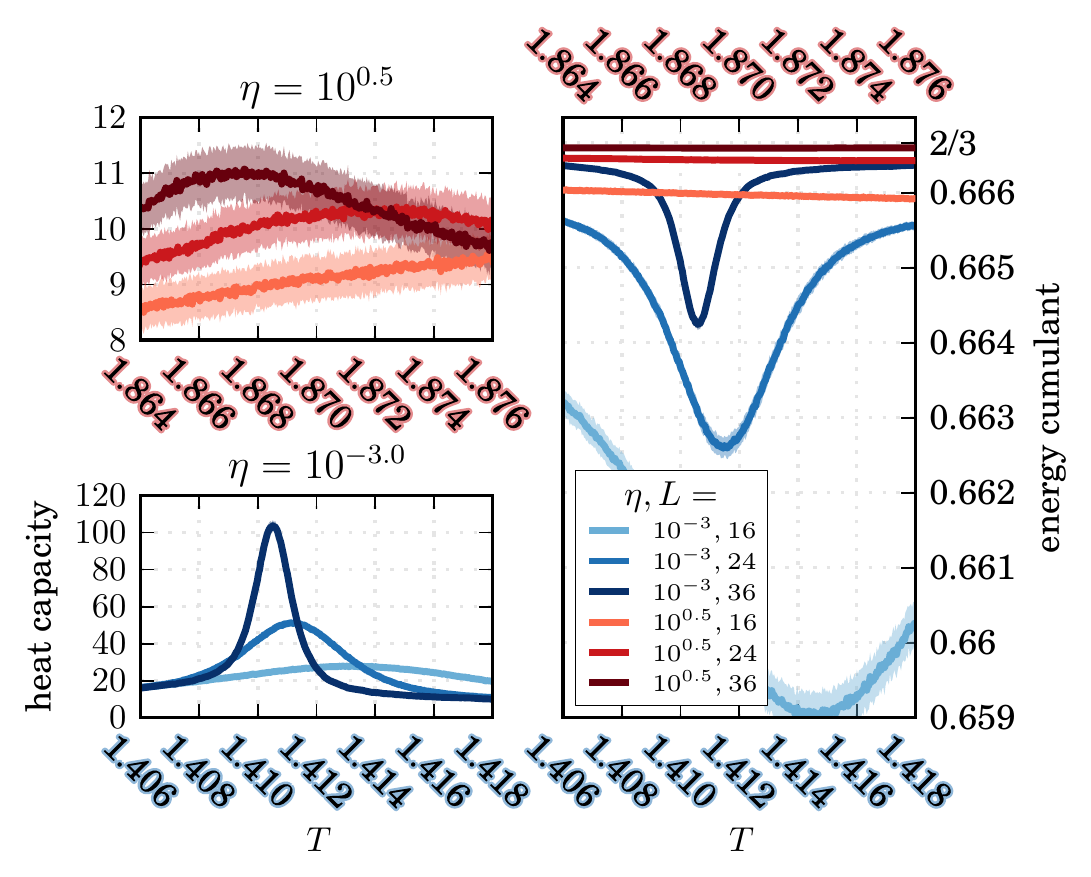}
	\caption{Scaling behavior of the heat capacity (left column) and energy cumulant (right) for $\eta=10\sp{-3}$ (with $\lambda<\xi_J$, blue lines) and $\eta=10\sp{0.5}$ (with $\lambda>\xi_J$, red lines), indicates a first-order and a continuous 
	phase transition respectively. Shaded regions around curves correspond to estimated errors.}
	\label{fig:enbinder}
\end{figure}

Consider next the scaling properties for the two-band model at the point $\eta=10^{0.5}$ which is in the region of $\lambda>\xi_J$ in Fig. \ref{fig:cmax_etavar} (with $\xi_J\approx 0.40$). In Fig. \ref{fig:enbinder} it is seen that the energy cumulant $V_L=1-\langle U\sp4\rangle/(3\langle U\sp2\rangle\sp2)$ has for $\eta=10^{-3}$ a minimum that is persistent against scaling, as it should for a first-order transition \cite{PhysRevB.34.1841}. For $\eta=10^{0.5}$ however, the energy cumulant has no distinct minimum, a behavior which is consistent with a continuous phase transition where the energy cumulant reaches the trivial limit of $2/3$ everywhere in the thermodynamic limit. Indeed, in the limit of strong interband coupling the system should recover the standard 3D inverted-XY phase transition like in a single-component model \cite{Peskin1978122,PhysRevLett.47.1556}.

We interpret the results as suggestive to an emergent attractive interaction between vortex lines.
It has been recently demonstrated that in a single-component directed-loops (${\bf j}$-current) model, a modification of the vortex interaction potential
from short-range repulsive to short-range non-monotonic asymptotically attractive,
causes a conversion of the inverted-XY transition to a first-order one \cite{meier2015fluctuation}.
The model under consideration here features fluctuation-induced attractive inter-vortex forces.
Indeed, if the Josephson coupling is set to zero, 
a fractional vortex has a logarithmically divergent energy
while co-directed  individual vortices interact 
logarithmically at large distances.
For a two-dimensional cross-section of a pair of fractional vortices in the two-component model with $|\psi_1 |^2=|\psi_2 |^2=|\psi|^2$ and phase winding
in the first condensate, the interaction energy  \cite{PhysRevLett.89.067001,babaev2004phase}
is:  
\begin{eqnarray}
E^{int}_{(1,0)+(1,0)}&=&\pi |\psi |^2\, \,
\log\frac{R}{r}
\, + \,   \pi\, |\psi |^2 K_0(r/\lambda).
 \label{fracint1}
\end{eqnarray}
The interaction between $(1,0)$- and $(0,1)$-vortices contains an attractive logarithmic part
and a repulsive exponentially screened part:
\begin{eqnarray}
E^{int}_{(1,0)+(0,1)}&=&-\pi |\psi |^2\, 
\log\frac{R}{r}\, + \,   {\pi} |\psi |^2   K_0(r/\lambda).
\label{fracint2}
\end{eqnarray}

In the absence of fluctuations, a composite vortex line is an axially symmetric object
with finite energy per unit length. The attractive and repulsive
forces cancel in the small separation limit \cite{PhysRevLett.89.067001,babaev2004phase},
however at separations larger than $\lambda$ a pair of co-directed fractional vortices
can be mapped onto an electric dipole.
Due to the long-range nature of the dipolar interaction, the thermal splitting 
of composite vortices into fractional vortices will lead to long-range attractive, short-range
repulsive interaction.
When a non-zero Josephson coupling is included, the interaction between fractional vortices
is changed to linearly attractive at distances larger than $\xi_J$ \cite{PhysRevLett.89.067001}.
Then the splitting fluctuations of fractional vortices are largely confined 
within the range $\xi_J$. The Josephson coupling also changes the dipolar interactions:
the phase difference mode becomes massive and dipolar forces become short-range
with the range again set by $\xi_J$. 

The results of our simulations suggest the scenario of a first-order phase transition originating in an attractive vortex interaction, i.e. in that case the superconducting phase transition beyond mean-field approximation is associated with proliferation of directed loops with induced dipolar attractive forces.
These forces should tend to induce formation of polarized vortex  clusters and phase separation.
In the three-component case elementary vortices can be mapped
onto Coulomb charges of different ``colors" \cite{smiseth2005field}. In that case the dipolar forces are stronger and
we observe a more pronounced first-order phase transition. On the other hand
we found no signatures of a first-order transition where the Josephson length is smaller than the 
magnetic field penetration length and the model can be approximated by 
single-species 
repulsively interacting directed loops. 
Note that a similar fluctuation-induced dipolar
interaction between topological defects exists in gauge theories
which break higher symmetry as well ($U(1)\times U(1)$ or $SU(2)$), where first-order phase transitions were also reported \cite{cond-mat/0501052,PhysRevLett.101.050405,chen2013deconfined,Kuklov20061602,PhysRevB.82.134511,
0805.2578, herland2013phase}. Indeed for the $U(1)\times U(1)$ 
models the composite vortices are the lowest-energy topological excitations
for any non-zero value of electric charge.
Also in the $SU(2)$ case one has composite vortices and Hopfions  \cite{babaev2001hidden,babaev2009non} which should have 
attractive interaction. If electric charge is decreased, the length scale 
at which the dipolar interaction sets in is increased, so in this scenario it requires
a larger system to detect first order phase transition at low electric charge.
Yet, the composite vortices are the lowest energy topological defects
at any non-zero value of electric charge and cannot be a priori neglected.  The energy of topological
defects is indeed different
in models that have global symmetry, such as  $U(1)\times U(1)$
which can have composite vortices due to dissipationless drag interaction.
There, in contrast to the gauge theories, despite the
existence of a paired phase  \cite{PhysRevLett.90.100401,PhysRevLett.92.050402},
there is indeed a tricritical point and a continuous phase transition 
at low inter component coupling \cite{PhysRevLett.92.030403,Kuklov20061602,PhysRevB.77.144519}.

In conclusion, we have reported that in the London limit $U(1)$ multiband superconductors can have a fluctuation-induced first-order phase transition in zero external field, in contrast to the inverted-XY phase transitions in single-band $U(1)$ London models \cite{Thomas1978513,Peskin1978122,PhysRevLett.47.1556}. We argued that the mechanism responsible for driving the phase transition to first-order is fluctuation-induced dipolar interactions between composite vortices. The mechanism should also apply for other multicomponent gauge theories including theories with higher broken symmetry.

\appendix 
\section{Details of the numerical discretization}
The system is discretized onto a square three-dimensional lattice 
$L\times L\times L$ 
with isotropic grid spacing 
$h$ 
and periodic boundary conditions imposed in all directions. Every site 
$i$ 
has the phases 
$\theta_i^{(1)}$, $\theta_i^{(2)}$ 
and the three components $A_{ix}$, $A_{iy}$ and $A_{iz}$ of the vector $(\mathbf{A})_i$. 
The phase gradients are discretized by the finite difference approximation 
$
[\nabla\theta\sp{(a)}]_{i\mu}
=(\theta^{(a)}_{i+\hat{\mu}}-\theta^{(a)}_{i})/h
$ 
with 
$\mu=x,y,z$, 
so that the kinetic energy density on site $i$ can be written 
$
\sum_\mu(\theta^{(a)}_{i+\hat{\mu}}
-\theta^{(a)}_i+hA_{i\mu})^2/h\sp2
$. 
The magnetic field energy density term on site $i$ is calculated from the definition of a curl as an infinitesimal circulation. Using the notation $x'=y$, $y'=z$ and $z'=x$, we can write the circulation around the plaquette $P$ with corner in $i$, with area $h$ and normal $\hat{\mu}$ as 
$
((\nabla\times\mathbf{A})\cdot\hat{\mathbf{\mu}})_i\approx
(h^{-2}\oint_{P}\mathbf{A}\cdot d\mathbf{r})_i=
h^{-2}
(hA_{i\mu'}+
hA_{i+\hat{\mu}',\mu''}
-hA_{i+\hat{\mu}'',\mu'}
-hA_{i\mu''})
$. Denoting
$
A_{i\mu'}+
A_{i+\hat{\mu}',\mu''}
-A_{i+\hat{\mu}'',\mu'}
-A_{i\mu''}
=	
[\Delta\times\mathbf{A}]_{i\mu}
$
we obtain
$
	\lambda^2\left((\nabla\times\mathbf{A})^2\right)_i
	=\lambda^2\sum_\mu
((\nabla\times\mathbf{A})\cdot\hat{\mathbf{\mu}})_i^2=
	(\lambda/h)^2\sum_\mu[\Delta\times 		h\mathbf{A}]_{i\mu}^2/h\sp2.
$ We may furthermore rewrite the Josephson term as
$
\xi_J^{-2}\cos(\theta^{(1)}-\theta^{(2)})
=(h/\xi_J)^2\cos(\theta^{(1)}-\theta^{(2)})/h\sp2
$. 

Since all terms in the Hamiltonian contains the factor $1/h\sp2$, both length scales appear in the units of $h$, and we can absorb $h$ into $\mathbf{A}$, we may set $h=1$ without loss of generality. After rewriting the kinetic energy terms into cosine $XY$-terms ($\cos(x)\approx1-x\sp2/2$) we obtain the Hamiltonian \eqref{eq:Hamiltonian}.

\section{Acknowledgements}
{

The work was supported by the Swedish Research Council grants 
642-2013-7837,  325-2009-7664.  The work was supported in part by the National Science Foundation
under Grant No. PHYS-1066293 and the hospitality of the Aspen Center
for Physics.}
The computations were performed on resources provided by the Swedish 
National Infrastructure for Computing (SNIC) at National Supercomputer Center 
at Link\"oping, Sweden.

\bibliographystyle{apsrev4-1}
\bibliography{references.bib}

\begin{thebibliography}{46}%
\makeatletter
\providecommand \@ifxundefined [1]{%
 \@ifx{#1\undefined}
}%
\providecommand \@ifnum [1]{%
 \ifnum #1\expandafter \@firstoftwo
 \else \expandafter \@secondoftwo
 \fi
}%
\providecommand \@ifx [1]{%
 \ifx #1\expandafter \@firstoftwo
 \else \expandafter \@secondoftwo
 \fi
}%
\providecommand \natexlab [1]{#1}%
\providecommand \enquote  [1]{``#1''}%
\providecommand \bibnamefont  [1]{#1}%
\providecommand \bibfnamefont [1]{#1}%
\providecommand \citenamefont [1]{#1}%
\providecommand \href@noop [0]{\@secondoftwo}%
\providecommand \href [0]{\begingroup \@sanitize@url \@href}%
\providecommand \@href[1]{\@@startlink{#1}\@@href}%
\providecommand \@@href[1]{\endgroup#1\@@endlink}%
\providecommand \@sanitize@url [0]{\catcode `\\12\catcode `\$12\catcode
  `\&12\catcode `\#12\catcode `\^12\catcode `\_12\catcode `\%12\relax}%
\providecommand \@@startlink[1]{}%
\providecommand \@@endlink[0]{}%
\providecommand \url  [0]{\begingroup\@sanitize@url \@url }%
\providecommand \@url [1]{\endgroup\@href {#1}{\urlprefix }}%
\providecommand \urlprefix  [0]{URL }%
\providecommand \Eprint [0]{\href }%
\providecommand \doibase [0]{http://dx.doi.org/}%
\providecommand \selectlanguage [0]{\@gobble}%
\providecommand \bibinfo  [0]{\@secondoftwo}%
\providecommand \bibfield  [0]{\@secondoftwo}%
\providecommand \translation [1]{[#1]}%
\providecommand \BibitemOpen [0]{}%
\providecommand \bibitemStop [0]{}%
\providecommand \bibitemNoStop [0]{.\EOS\space}%
\providecommand \EOS [0]{\spacefactor3000\relax}%
\providecommand \BibitemShut  [1]{\csname bibitem#1\endcsname}%
\let\auto@bib@innerbib\@empty
\bibitem [{\citenamefont {Halperin}\ \emph {et~al.}(1974)\citenamefont
  {Halperin}, \citenamefont {Lubensky},\ and\ \citenamefont
  {Ma}}]{PhysRevLett.32.292}%
  \BibitemOpen
  \bibfield  {author} {\bibinfo {author} {\bibfnamefont {B.~I.}\ \bibnamefont
  {Halperin}}, \bibinfo {author} {\bibfnamefont {T.~C.}\ \bibnamefont
  {Lubensky}}, \ and\ \bibinfo {author} {\bibfnamefont {S.-k.}\ \bibnamefont
  {Ma}},\ }\href {\doibase 10.1103/PhysRevLett.32.292} {\bibfield  {journal}
  {\bibinfo  {journal} {Phys. Rev. Lett.}\ }\textbf {\bibinfo {volume} {32}},\
  \bibinfo {pages} {292} (\bibinfo {year} {1974})}\BibitemShut {NoStop}%
\bibitem [{\citenamefont {Coleman}\ and\ \citenamefont
  {Weinberg}(1973)}]{PhysRevD.7.1888}%
  \BibitemOpen
  \bibfield  {author} {\bibinfo {author} {\bibfnamefont {S.}~\bibnamefont
  {Coleman}}\ and\ \bibinfo {author} {\bibfnamefont {E.}~\bibnamefont
  {Weinberg}},\ }\href {\doibase 10.1103/PhysRevD.7.1888} {\bibfield  {journal}
  {\bibinfo  {journal} {Phys. Rev. D}\ }\textbf {\bibinfo {volume} {7}},\
  \bibinfo {pages} {1888} (\bibinfo {year} {1973})}\BibitemShut {NoStop}%
\bibitem [{\citenamefont {Thomas}\ and\ \citenamefont
  {Stone}(1978)}]{Thomas1978513}%
  \BibitemOpen
  \bibfield  {author} {\bibinfo {author} {\bibfnamefont {P.~R.}\ \bibnamefont
  {Thomas}}\ and\ \bibinfo {author} {\bibfnamefont {M.}~\bibnamefont {Stone}},\
  }\href {\doibase http://dx.doi.org/10.1016/0550-3213(78)90383-8} {\bibfield
  {journal} {\bibinfo  {journal} {Nuclear Physics B}\ }\textbf {\bibinfo
  {volume} {144}},\ \bibinfo {pages} {513 } (\bibinfo {year}
  {1978})}\BibitemShut {NoStop}%
\bibitem [{\citenamefont {Peskin}(1978)}]{Peskin1978122}%
  \BibitemOpen
  \bibfield  {author} {\bibinfo {author} {\bibfnamefont {M.~E.}\ \bibnamefont
  {Peskin}},\ }\href {\doibase http://dx.doi.org/10.1016/0003-4916(78)90252-X}
  {\bibfield  {journal} {\bibinfo  {journal} {Annals of Physics}\ }\textbf
  {\bibinfo {volume} {113}},\ \bibinfo {pages} {122 } (\bibinfo {year}
  {1978})}\BibitemShut {NoStop}%
\bibitem [{\citenamefont {Dasgupta}\ and\ \citenamefont
  {Halperin}(1981)}]{PhysRevLett.47.1556}%
  \BibitemOpen
  \bibfield  {author} {\bibinfo {author} {\bibfnamefont {C.}~\bibnamefont
  {Dasgupta}}\ and\ \bibinfo {author} {\bibfnamefont {B.}~\bibnamefont
  {Halperin}},\ }\href {\doibase 10.1103/PhysRevLett.47.1556} {\bibfield
  {journal} {\bibinfo  {journal} {Phys. Rev. Lett.}\ }\textbf {\bibinfo
  {volume} {47}},\ \bibinfo {pages} {1556} (\bibinfo {year}
  {1981})}\BibitemShut {NoStop}%
\bibitem [{\citenamefont {Olsson}\ and\ \citenamefont
  {Teitel}(1998)}]{PhysRevLett.80.1964}%
  \BibitemOpen
  \bibfield  {author} {\bibinfo {author} {\bibfnamefont {P.}~\bibnamefont
  {Olsson}}\ and\ \bibinfo {author} {\bibfnamefont {S.}~\bibnamefont
  {Teitel}},\ }\href {\doibase 10.1103/PhysRevLett.80.1964} {\bibfield
  {journal} {\bibinfo  {journal} {Phys. Rev. Lett.}\ }\textbf {\bibinfo
  {volume} {80}},\ \bibinfo {pages} {1964} (\bibinfo {year}
  {1998})}\BibitemShut {NoStop}%
\bibitem [{\citenamefont {Kleinert}(1982)}]{kleinert1982disorder}%
  \BibitemOpen
  \bibfield  {author} {\bibinfo {author} {\bibfnamefont {H.}~\bibnamefont
  {Kleinert}},\ }\href@noop {} {\bibfield  {journal} {\bibinfo  {journal}
  {Lettere Al Nuovo Cimento Series 2}\ }\textbf {\bibinfo {volume} {35}},\
  \bibinfo {pages} {405} (\bibinfo {year} {1982})}\BibitemShut {NoStop}%
\bibitem [{\citenamefont {Bartholomew}(1983)}]{bartholomew1983phase}%
  \BibitemOpen
  \bibfield  {author} {\bibinfo {author} {\bibfnamefont {J.}~\bibnamefont
  {Bartholomew}},\ }\href@noop {} {\bibfield  {journal} {\bibinfo  {journal}
  {Physical Review B}\ }\textbf {\bibinfo {volume} {28}},\ \bibinfo {pages}
  {5378} (\bibinfo {year} {1983})}\BibitemShut {NoStop}%
\bibitem [{\citenamefont {Herbut}\ and\ \citenamefont {Te\ifmmode
  \check{s}\else \v{s}\fi{}anovi\ifmmode~\acute{c}\else
  \'{c}\fi{}}(1996)}]{PhysRevLett.76.4588}%
  \BibitemOpen
  \bibfield  {author} {\bibinfo {author} {\bibfnamefont {I.~F.}\ \bibnamefont
  {Herbut}}\ and\ \bibinfo {author} {\bibfnamefont {Z.}~\bibnamefont
  {Te\ifmmode \check{s}\else \v{s}\fi{}anovi\ifmmode~\acute{c}\else
  \'{c}\fi{}}},\ }\href {\doibase 10.1103/PhysRevLett.76.4588} {\bibfield
  {journal} {\bibinfo  {journal} {Phys. Rev. Lett.}\ }\textbf {\bibinfo
  {volume} {76}},\ \bibinfo {pages} {4588} (\bibinfo {year}
  {1996})}\BibitemShut {NoStop}%
\bibitem [{\citenamefont {Mo}\ \emph {et~al.}(2002)\citenamefont {Mo},
  \citenamefont {Hove},\ and\ \citenamefont {Sudb\o{}}}]{PhysRevB.65.104501}%
  \BibitemOpen
  \bibfield  {author} {\bibinfo {author} {\bibfnamefont {S.}~\bibnamefont
  {Mo}}, \bibinfo {author} {\bibfnamefont {J.}~\bibnamefont {Hove}}, \ and\
  \bibinfo {author} {\bibfnamefont {A.}~\bibnamefont {Sudb\o{}}},\ }\href
  {\doibase 10.1103/PhysRevB.65.104501} {\bibfield  {journal} {\bibinfo
  {journal} {Phys. Rev. B}\ }\textbf {\bibinfo {volume} {65}},\ \bibinfo
  {pages} {104501} (\bibinfo {year} {2002})}\BibitemShut {NoStop}%
\bibitem [{\citenamefont {{Babaev}}(2004)}]{babaev2004phase}%
  \BibitemOpen
  \bibfield  {author} {\bibinfo {author} {\bibfnamefont {E.}~\bibnamefont
  {{Babaev}}},\ }\href@noop {} {\enquote {\bibinfo {title} {{Phase diagram of
  planar U(1)${\times}$ U(1) superconductor . Condensation of vortices with
  fractional flux and a superfluid state}},}\ } (\bibinfo {year} {2004}),\
  \Eprint {http://arxiv.org/abs/cond-mat/0201547} {cond-mat/0201547}
  \BibitemShut {NoStop}%
\bibitem [{\citenamefont {Babaev}\ \emph {et~al.}(2004)\citenamefont {Babaev},
  \citenamefont {Sudb{\o}},\ and\ \citenamefont
  {Ashcroft}}]{babaev2004superconductor}%
  \BibitemOpen
  \bibfield  {author} {\bibinfo {author} {\bibfnamefont {E.}~\bibnamefont
  {Babaev}}, \bibinfo {author} {\bibfnamefont {A.}~\bibnamefont {Sudb{\o}}}, \
  and\ \bibinfo {author} {\bibfnamefont {N.}~\bibnamefont {Ashcroft}},\
  }\href@noop {} {\bibfield  {journal} {\bibinfo  {journal} {Nature}\ }\textbf
  {\bibinfo {volume} {431}},\ \bibinfo {pages} {666} (\bibinfo {year}
  {2004})}\BibitemShut {NoStop}%
\bibitem [{\citenamefont {Herland}\ \emph {et~al.}(2010)\citenamefont
  {Herland}, \citenamefont {Babaev},\ and\ \citenamefont
  {Sudb\o{}}}]{PhysRevB.82.134511}%
  \BibitemOpen
  \bibfield  {author} {\bibinfo {author} {\bibfnamefont {E.}~\bibnamefont
  {Herland}}, \bibinfo {author} {\bibfnamefont {E.}~\bibnamefont {Babaev}}, \
  and\ \bibinfo {author} {\bibfnamefont {A.}~\bibnamefont {Sudb\o{}}},\ }\href
  {\doibase 10.1103/PhysRevB.82.134511} {\bibfield  {journal} {\bibinfo
  {journal} {Phys. Rev. B}\ }\textbf {\bibinfo {volume} {82}},\ \bibinfo
  {pages} {134511} (\bibinfo {year} {2010})}\BibitemShut {NoStop}%
\bibitem [{\citenamefont {Motrunich}\ and\ \citenamefont
  {Vishwanath}(2004)}]{PhysRevB.70.075104}%
  \BibitemOpen
  \bibfield  {author} {\bibinfo {author} {\bibfnamefont {O.~I.}\ \bibnamefont
  {Motrunich}}\ and\ \bibinfo {author} {\bibfnamefont {A.}~\bibnamefont
  {Vishwanath}},\ }\href {\doibase 10.1103/PhysRevB.70.075104} {\bibfield
  {journal} {\bibinfo  {journal} {Phys. Rev. B}\ }\textbf {\bibinfo {volume}
  {70}},\ \bibinfo {pages} {075104} (\bibinfo {year} {2004})}\BibitemShut
  {NoStop}%
\bibitem [{\citenamefont {Senthil}\ \emph {et~al.}(2004)\citenamefont
  {Senthil}, \citenamefont {Vishwanath}, \citenamefont {Balents}, \citenamefont
  {Sachdev},\ and\ \citenamefont {Fisher}}]{Senthil05032004}%
  \BibitemOpen
  \bibfield  {author} {\bibinfo {author} {\bibfnamefont {T.}~\bibnamefont
  {Senthil}}, \bibinfo {author} {\bibfnamefont {A.}~\bibnamefont {Vishwanath}},
  \bibinfo {author} {\bibfnamefont {L.}~\bibnamefont {Balents}}, \bibinfo
  {author} {\bibfnamefont {S.}~\bibnamefont {Sachdev}}, \ and\ \bibinfo
  {author} {\bibfnamefont {M.~P.~A.}\ \bibnamefont {Fisher}},\ }\href {\doibase
  10.1126/science.1091806} {\bibfield  {journal} {\bibinfo  {journal}
  {Science}\ }\textbf {\bibinfo {volume} {303}},\ \bibinfo {pages} {1490}
  (\bibinfo {year} {2004})},\ \Eprint
  {http://arxiv.org/abs/http://www.sciencemag.org/content/303/5663/1490.full.pdf}
  {http://www.sciencemag.org/content/303/5663/1490.full.pdf} \BibitemShut
  {NoStop}%
\bibitem [{\citenamefont {Sachdev}(2008)}]{sachdev2008quantum}%
  \BibitemOpen
  \bibfield  {author} {\bibinfo {author} {\bibfnamefont {S.}~\bibnamefont
  {Sachdev}},\ }\href@noop {} {\bibfield  {journal} {\bibinfo  {journal}
  {Nature Physics}\ }\textbf {\bibinfo {volume} {4}},\ \bibinfo {pages} {173}
  (\bibinfo {year} {2008})}\BibitemShut {NoStop}%
\bibitem [{\citenamefont {Chen}\ \emph {et~al.}(2013)\citenamefont {Chen},
  \citenamefont {Huang}, \citenamefont {Deng}, \citenamefont {Kuklov},
  \citenamefont {Prokof'ev},\ and\ \citenamefont
  {Svistunov}}]{chen2013deconfined}%
  \BibitemOpen
  \bibfield  {author} {\bibinfo {author} {\bibfnamefont {K.}~\bibnamefont
  {Chen}}, \bibinfo {author} {\bibfnamefont {Y.}~\bibnamefont {Huang}},
  \bibinfo {author} {\bibfnamefont {Y.}~\bibnamefont {Deng}}, \bibinfo {author}
  {\bibfnamefont {A.}~\bibnamefont {Kuklov}}, \bibinfo {author} {\bibfnamefont
  {N.}~\bibnamefont {Prokof'ev}}, \ and\ \bibinfo {author} {\bibfnamefont
  {B.}~\bibnamefont {Svistunov}},\ }\href@noop {} {\bibfield  {journal}
  {\bibinfo  {journal} {Physical review letters}\ }\textbf {\bibinfo {volume}
  {110}},\ \bibinfo {pages} {185701} (\bibinfo {year} {2013})}\BibitemShut
  {NoStop}%
\bibitem [{\citenamefont {Motrunich}\ and\ \citenamefont
  {Vishwanath}(2008)}]{motrunich2008comparative}%
  \BibitemOpen
  \bibfield  {author} {\bibinfo {author} {\bibfnamefont {O.~I.}\ \bibnamefont
  {Motrunich}}\ and\ \bibinfo {author} {\bibfnamefont {A.}~\bibnamefont
  {Vishwanath}},\ }\href@noop {} {\bibfield  {journal} {\bibinfo  {journal}
  {arXiv preprint arXiv:0805.1494}\ } (\bibinfo {year} {2008})}\BibitemShut
  {NoStop}%
\bibitem [{\citenamefont {Kuklov}\ \emph {et~al.}(2005)\citenamefont {Kuklov},
  \citenamefont {Prokof'ev},\ and\ \citenamefont
  {Svistunov}}]{cond-mat/0501052}%
  \BibitemOpen
  \bibfield  {author} {\bibinfo {author} {\bibfnamefont {A.}~\bibnamefont
  {Kuklov}}, \bibinfo {author} {\bibfnamefont {N.}~\bibnamefont {Prokof'ev}}, \
  and\ \bibinfo {author} {\bibfnamefont {B.}~\bibnamefont {Svistunov}},\
  }\href@noop {} {\enquote {\bibinfo {title} {Ginzburg-landau-wilson aspect of
  deconfined criticality},}\ } (\bibinfo {year} {2005}),\ \Eprint
  {http://arxiv.org/abs/arXiv:cond-mat/0501052} {arXiv:cond-mat/0501052}
  \BibitemShut {NoStop}%
\bibitem [{\citenamefont {Kuklov}\ \emph
  {et~al.}(2008{\natexlab{a}})\citenamefont {Kuklov}, \citenamefont
  {Matsumoto}, \citenamefont {Prokof'ev}, \citenamefont {Svistunov},\ and\
  \citenamefont {Troyer}}]{PhysRevLett.101.050405}%
  \BibitemOpen
  \bibfield  {author} {\bibinfo {author} {\bibfnamefont {A.}~\bibnamefont
  {Kuklov}}, \bibinfo {author} {\bibfnamefont {M.}~\bibnamefont {Matsumoto}},
  \bibinfo {author} {\bibfnamefont {N.}~\bibnamefont {Prokof'ev}}, \bibinfo
  {author} {\bibfnamefont {B.}~\bibnamefont {Svistunov}}, \ and\ \bibinfo
  {author} {\bibfnamefont {M.}~\bibnamefont {Troyer}},\ }\href {\doibase
  10.1103/PhysRevLett.101.050405} {\bibfield  {journal} {\bibinfo  {journal}
  {Phys. Rev. Lett.}\ }\textbf {\bibinfo {volume} {101}},\ \bibinfo {pages}
  {050405} (\bibinfo {year} {2008}{\natexlab{a}})}\BibitemShut {NoStop}%
\bibitem [{\citenamefont {Kuklov}\ \emph {et~al.}(2006)\citenamefont {Kuklov},
  \citenamefont {Prokof'ev}, \citenamefont {Svistunov},\ and\ \citenamefont
  {Troyer}}]{Kuklov20061602}%
  \BibitemOpen
  \bibfield  {author} {\bibinfo {author} {\bibfnamefont {A.}~\bibnamefont
  {Kuklov}}, \bibinfo {author} {\bibfnamefont {N.}~\bibnamefont {Prokof'ev}},
  \bibinfo {author} {\bibfnamefont {B.}~\bibnamefont {Svistunov}}, \ and\
  \bibinfo {author} {\bibfnamefont {M.}~\bibnamefont {Troyer}},\ }\href
  {\doibase http://dx.doi.org/10.1016/j.aop.2006.04.007} {\bibfield  {journal}
  {\bibinfo  {journal} {Annals of Physics}\ }\textbf {\bibinfo {volume}
  {321}},\ \bibinfo {pages} {1602 } (\bibinfo {year} {2006})},\ \bibinfo {note}
  {july 2006 Special Issue}\BibitemShut {NoStop}%
\bibitem [{\citenamefont {Kuklov}\ \emph
  {et~al.}(2008{\natexlab{b}})\citenamefont {Kuklov}, \citenamefont
  {Matsumoto}, \citenamefont {Prokof'ev}, \citenamefont {Svistunov},\ and\
  \citenamefont {Troyer}}]{0805.2578}%
  \BibitemOpen
  \bibfield  {author} {\bibinfo {author} {\bibfnamefont {A.}~\bibnamefont
  {Kuklov}}, \bibinfo {author} {\bibfnamefont {M.}~\bibnamefont {Matsumoto}},
  \bibinfo {author} {\bibfnamefont {N.}~\bibnamefont {Prokof'ev}}, \bibinfo
  {author} {\bibfnamefont {B.}~\bibnamefont {Svistunov}}, \ and\ \bibinfo
  {author} {\bibfnamefont {M.}~\bibnamefont {Troyer}},\ }\href@noop {}
  {\enquote {\bibinfo {title} {Comment on ``comparative study of higgs
  transition in one-component and two-component lattice superconductor
  models"},}\ } (\bibinfo {year} {2008}{\natexlab{b}}),\ \Eprint
  {http://arxiv.org/abs/arXiv:0805.2578} {arXiv:0805.2578} \BibitemShut
  {NoStop}%
\bibitem [{\citenamefont {{Kragset}}\ \emph {et~al.}(2006)\citenamefont
  {{Kragset}}, \citenamefont {{Sm{\o}rgrav}}, \citenamefont {{Hove}},
  \citenamefont {{Nogueira}},\ and\ \citenamefont
  {{Sudb{\o}}}}]{2006PhRvL..97x7201K}%
  \BibitemOpen
  \bibfield  {author} {\bibinfo {author} {\bibfnamefont {S.}~\bibnamefont
  {{Kragset}}}, \bibinfo {author} {\bibfnamefont {E.}~\bibnamefont
  {{Sm{\o}rgrav}}}, \bibinfo {author} {\bibfnamefont {J.}~\bibnamefont
  {{Hove}}}, \bibinfo {author} {\bibfnamefont {F.~S.}\ \bibnamefont
  {{Nogueira}}}, \ and\ \bibinfo {author} {\bibfnamefont {A.}~\bibnamefont
  {{Sudb{\o}}}},\ }\href {\doibase 10.1103/PhysRevLett.97.247201} {\bibfield
  {journal} {\bibinfo  {journal} {Physical Review Letters}\ }\textbf {\bibinfo
  {volume} {97}},\ \bibinfo {eid} {247201} (\bibinfo {year} {2006})},\ \Eprint
  {http://arxiv.org/abs/cond-mat/0609336} {cond-mat/0609336} \BibitemShut
  {NoStop}%
\bibitem [{\citenamefont {Herland}\ \emph {et~al.}(2013)\citenamefont
  {Herland}, \citenamefont {Bojesen}, \citenamefont {Babaev},\ and\
  \citenamefont {Sudb{\o}}}]{herland2013phase}%
  \BibitemOpen
  \bibfield  {author} {\bibinfo {author} {\bibfnamefont {E.~V.}\ \bibnamefont
  {Herland}}, \bibinfo {author} {\bibfnamefont {T.~A.}\ \bibnamefont
  {Bojesen}}, \bibinfo {author} {\bibfnamefont {E.}~\bibnamefont {Babaev}}, \
  and\ \bibinfo {author} {\bibfnamefont {A.}~\bibnamefont {Sudb{\o}}},\
  }\href@noop {} {\bibfield  {journal} {\bibinfo  {journal} {Physical Review
  B}\ }\textbf {\bibinfo {volume} {87}},\ \bibinfo {pages} {134503} (\bibinfo
  {year} {2013})}\BibitemShut {NoStop}%
\bibitem [{\citenamefont {Bojesen}\ \emph {et~al.}(2014)\citenamefont
  {Bojesen}, \citenamefont {Babaev},\ and\ \citenamefont
  {Sudb\o{}}}]{PhysRevB.89.104509}%
  \BibitemOpen
  \bibfield  {author} {\bibinfo {author} {\bibfnamefont {T.~A.}\ \bibnamefont
  {Bojesen}}, \bibinfo {author} {\bibfnamefont {E.}~\bibnamefont {Babaev}}, \
  and\ \bibinfo {author} {\bibfnamefont {A.}~\bibnamefont {Sudb\o{}}},\ }\href
  {\doibase 10.1103/PhysRevB.89.104509} {\bibfield  {journal} {\bibinfo
  {journal} {Phys. Rev. B}\ }\textbf {\bibinfo {volume} {89}},\ \bibinfo
  {pages} {104509} (\bibinfo {year} {2014})}\BibitemShut {NoStop}%
\bibitem [{\citenamefont {Suhl}\ \emph {et~al.}(1959)\citenamefont {Suhl},
  \citenamefont {Matthias},\ and\ \citenamefont {Walker}}]{PhysRevLett.3.552}%
  \BibitemOpen
  \bibfield  {author} {\bibinfo {author} {\bibfnamefont {H.}~\bibnamefont
  {Suhl}}, \bibinfo {author} {\bibfnamefont {B.}~\bibnamefont {Matthias}}, \
  and\ \bibinfo {author} {\bibfnamefont {L.}~\bibnamefont {Walker}},\ }\href
  {\doibase 10.1103/PhysRevLett.3.552} {\bibfield  {journal} {\bibinfo
  {journal} {Phys. Rev. Lett.}\ }\textbf {\bibinfo {volume} {3}},\ \bibinfo
  {pages} {552} (\bibinfo {year} {1959})}\BibitemShut {NoStop}%
\bibitem [{\citenamefont {Moskalenko}(1959)}]{moskalenko1959superconductivity}%
  \BibitemOpen
  \bibfield  {author} {\bibinfo {author} {\bibfnamefont {V.}~\bibnamefont
  {Moskalenko}},\ }\href@noop {} {\bibfield  {journal} {\bibinfo  {journal}
  {Fiz. Metal. Metalloved}\ }\textbf {\bibinfo {volume} {8}},\ \bibinfo {pages}
  {2518} (\bibinfo {year} {1959})}\BibitemShut {NoStop}%
\bibitem [{\citenamefont {Leggett}(1966)}]{leggett1966number}%
  \BibitemOpen
  \bibfield  {author} {\bibinfo {author} {\bibfnamefont {A.}~\bibnamefont
  {Leggett}},\ }\href@noop {} {\bibfield  {journal} {\bibinfo  {journal}
  {Progress of Theoretical Physics}\ }\textbf {\bibinfo {volume} {36}},\
  \bibinfo {pages} {901} (\bibinfo {year} {1966})}\BibitemShut {NoStop}%
\bibitem [{\citenamefont {Smiseth}\ \emph {et~al.}(2005)\citenamefont
  {Smiseth}, \citenamefont {Sm{\o}rgrav}, \citenamefont {Babaev},\ and\
  \citenamefont {Sudb{\o}}}]{smiseth2005field}%
  \BibitemOpen
  \bibfield  {author} {\bibinfo {author} {\bibfnamefont {J.}~\bibnamefont
  {Smiseth}}, \bibinfo {author} {\bibfnamefont {E.}~\bibnamefont
  {Sm{\o}rgrav}}, \bibinfo {author} {\bibfnamefont {E.}~\bibnamefont {Babaev}},
  \ and\ \bibinfo {author} {\bibfnamefont {A.}~\bibnamefont {Sudb{\o}}},\
  }\href@noop {} {\bibfield  {journal} {\bibinfo  {journal} {Physical Review
  B}\ }\textbf {\bibinfo {volume} {71}},\ \bibinfo {pages} {214509} (\bibinfo
  {year} {2005})}\BibitemShut {NoStop}%
\bibitem [{\citenamefont {Meier}\ \emph {et~al.}(2015)\citenamefont {Meier},
  \citenamefont {Babaev},\ and\ \citenamefont {Wallin}}]{meier2015fluctuation}%
  \BibitemOpen
  \bibfield  {author} {\bibinfo {author} {\bibfnamefont {H.}~\bibnamefont
  {Meier}}, \bibinfo {author} {\bibfnamefont {E.}~\bibnamefont {Babaev}}, \
  and\ \bibinfo {author} {\bibfnamefont {M.}~\bibnamefont {Wallin}},\
  }\href@noop {} {\bibfield  {journal} {\bibinfo  {journal} {Physical Review
  B}\ }\textbf {\bibinfo {volume} {91}},\ \bibinfo {pages} {094508} (\bibinfo
  {year} {2015})}\BibitemShut {NoStop}%
\bibitem [{\citenamefont {Sm{\o}rgrav}\ \emph
  {et~al.}(2005{\natexlab{a}})\citenamefont {Sm{\o}rgrav}, \citenamefont
  {Babaev}, \citenamefont {Smiseth},\ and\ \citenamefont
  {Sudb{\o}}}]{smorgrav2005observation}%
  \BibitemOpen
  \bibfield  {author} {\bibinfo {author} {\bibfnamefont {E.}~\bibnamefont
  {Sm{\o}rgrav}}, \bibinfo {author} {\bibfnamefont {E.}~\bibnamefont {Babaev}},
  \bibinfo {author} {\bibfnamefont {J.}~\bibnamefont {Smiseth}}, \ and\
  \bibinfo {author} {\bibfnamefont {A.}~\bibnamefont {Sudb{\o}}},\ }\href@noop
  {} {\bibfield  {journal} {\bibinfo  {journal} {Physical review letters}\
  }\textbf {\bibinfo {volume} {95}},\ \bibinfo {pages} {135301} (\bibinfo
  {year} {2005}{\natexlab{a}})}\BibitemShut {NoStop}%
\bibitem [{\citenamefont {Sm{\o}rgrav}\ \emph
  {et~al.}(2005{\natexlab{b}})\citenamefont {Sm{\o}rgrav}, \citenamefont
  {Smiseth}, \citenamefont {Babaev},\ and\ \citenamefont
  {Sudb{\o}}}]{smorgrav2005vortex}%
  \BibitemOpen
  \bibfield  {author} {\bibinfo {author} {\bibfnamefont {E.}~\bibnamefont
  {Sm{\o}rgrav}}, \bibinfo {author} {\bibfnamefont {J.}~\bibnamefont
  {Smiseth}}, \bibinfo {author} {\bibfnamefont {E.}~\bibnamefont {Babaev}}, \
  and\ \bibinfo {author} {\bibfnamefont {A.}~\bibnamefont {Sudb{\o}}},\
  }\href@noop {} {\bibfield  {journal} {\bibinfo  {journal} {Physical review
  letters}\ }\textbf {\bibinfo {volume} {94}},\ \bibinfo {pages} {96401}
  (\bibinfo {year} {2005}{\natexlab{b}})}\BibitemShut {NoStop}%
\bibitem [{\citenamefont {Babaev}(2002)}]{PhysRevLett.89.067001}%
  \BibitemOpen
  \bibfield  {author} {\bibinfo {author} {\bibfnamefont {E.}~\bibnamefont
  {Babaev}},\ }\href {\doibase 10.1103/PhysRevLett.89.067001} {\bibfield
  {journal} {\bibinfo  {journal} {Phys. Rev. Lett.}\ }\textbf {\bibinfo
  {volume} {89}},\ \bibinfo {pages} {067001} (\bibinfo {year}
  {2002})}\BibitemShut {NoStop}%
\bibitem [{\citenamefont {Swendsen}\ and\ \citenamefont
  {Wang}(1986)}]{PhysRevLett.57.2607}%
  \BibitemOpen
  \bibfield  {author} {\bibinfo {author} {\bibfnamefont {R.~H.}\ \bibnamefont
  {Swendsen}}\ and\ \bibinfo {author} {\bibfnamefont {J.-S.}\ \bibnamefont
  {Wang}},\ }\href {\doibase 10.1103/PhysRevLett.57.2607} {\bibfield  {journal}
  {\bibinfo  {journal} {Phys. Rev. Lett.}\ }\textbf {\bibinfo {volume} {57}},\
  \bibinfo {pages} {2607} (\bibinfo {year} {1986})}\BibitemShut {NoStop}%
\bibitem [{\citenamefont {Earl}\ and\ \citenamefont {Deem}(2005)}]{B509983H}%
  \BibitemOpen
  \bibfield  {author} {\bibinfo {author} {\bibfnamefont {D.~J.}\ \bibnamefont
  {Earl}}\ and\ \bibinfo {author} {\bibfnamefont {M.~W.}\ \bibnamefont
  {Deem}},\ }\href {\doibase 10.1039/B509983H} {\bibfield  {journal} {\bibinfo
  {journal} {Phys. Chem. Chem. Phys.}\ }\textbf {\bibinfo {volume} {7}},\
  \bibinfo {pages} {3910} (\bibinfo {year} {2005})}\BibitemShut {NoStop}%
\bibitem [{\citenamefont {Newman}\ and\ \citenamefont
  {Barkema}(1999)}]{newman1999monte}%
  \BibitemOpen
  \bibfield  {author} {\bibinfo {author} {\bibfnamefont {E.}~\bibnamefont
  {Newman}}\ and\ \bibinfo {author} {\bibfnamefont {G.}~\bibnamefont
  {Barkema}},\ }\href {http://books.google.de/books?id=J5aLdDN4uFwC} {\emph
  {\bibinfo {title} {Monte Carlo Methods in Statistical Physics}}}\ (\bibinfo
  {publisher} {Clarendon Press},\ \bibinfo {year} {1999})\BibitemShut {NoStop}%
\bibitem [{\citenamefont {Manousiouthakis}\ and\ \citenamefont
  {Deem}(1999)}]{1.477973}%
  \BibitemOpen
  \bibfield  {author} {\bibinfo {author} {\bibfnamefont {V.~I.}\ \bibnamefont
  {Manousiouthakis}}\ and\ \bibinfo {author} {\bibfnamefont {M.~W.}\
  \bibnamefont {Deem}},\ }\href@noop {} {\bibfield  {journal} {\bibinfo
  {journal} {The Journal of Chemical Physics}\ }\textbf {\bibinfo {volume}
  {110}} (\bibinfo {year} {1999})}\BibitemShut {NoStop}%
\bibitem [{\citenamefont {Dahl}\ \emph {et~al.}(2008)\citenamefont {Dahl},
  \citenamefont {Babaev}, \citenamefont {Kragset},\ and\ \citenamefont
  {Sudb\o{}}}]{PhysRevB.77.144519}%
  \BibitemOpen
  \bibfield  {author} {\bibinfo {author} {\bibfnamefont {E.}~\bibnamefont
  {Dahl}}, \bibinfo {author} {\bibfnamefont {E.}~\bibnamefont {Babaev}},
  \bibinfo {author} {\bibfnamefont {S.}~\bibnamefont {Kragset}}, \ and\
  \bibinfo {author} {\bibfnamefont {A.}~\bibnamefont {Sudb\o{}}},\ }\href
  {\doibase 10.1103/PhysRevB.77.144519} {\bibfield  {journal} {\bibinfo
  {journal} {Phys. Rev. B}\ }\textbf {\bibinfo {volume} {77}},\ \bibinfo
  {pages} {144519} (\bibinfo {year} {2008})}\BibitemShut {NoStop}%
\bibitem [{\citenamefont {Lee}\ and\ \citenamefont
  {Kosterlitz}(1990)}]{PhysRevLett.65.137}%
  \BibitemOpen
  \bibfield  {author} {\bibinfo {author} {\bibfnamefont {J.}~\bibnamefont
  {Lee}}\ and\ \bibinfo {author} {\bibfnamefont {J.}~\bibnamefont
  {Kosterlitz}},\ }\href {\doibase 10.1103/PhysRevLett.65.137} {\bibfield
  {journal} {\bibinfo  {journal} {Phys. Rev. Lett.}\ }\textbf {\bibinfo
  {volume} {65}},\ \bibinfo {pages} {137} (\bibinfo {year} {1990})}\BibitemShut
  {NoStop}%
\bibitem [{\citenamefont {Lee}\ and\ \citenamefont
  {Kosterlitz}(1991)}]{PhysRevB.43.3265}%
  \BibitemOpen
  \bibfield  {author} {\bibinfo {author} {\bibfnamefont {J.}~\bibnamefont
  {Lee}}\ and\ \bibinfo {author} {\bibfnamefont {J.}~\bibnamefont
  {Kosterlitz}},\ }\href {\doibase 10.1103/PhysRevB.43.3265} {\bibfield
  {journal} {\bibinfo  {journal} {Phys. Rev. B}\ }\textbf {\bibinfo {volume}
  {43}},\ \bibinfo {pages} {3265} (\bibinfo {year} {1991})}\BibitemShut
  {NoStop}%
\bibitem [{\citenamefont {Challa}\ \emph {et~al.}(1986)\citenamefont {Challa},
  \citenamefont {Landau},\ and\ \citenamefont {Binder}}]{PhysRevB.34.1841}%
  \BibitemOpen
  \bibfield  {author} {\bibinfo {author} {\bibfnamefont {M.}~\bibnamefont
  {Challa}}, \bibinfo {author} {\bibfnamefont {D.}~\bibnamefont {Landau}}, \
  and\ \bibinfo {author} {\bibfnamefont {K.}~\bibnamefont {Binder}},\ }\href
  {\doibase 10.1103/PhysRevB.34.1841} {\bibfield  {journal} {\bibinfo
  {journal} {Phys. Rev. B}\ }\textbf {\bibinfo {volume} {34}},\ \bibinfo
  {pages} {1841} (\bibinfo {year} {1986})}\BibitemShut {NoStop}%
\bibitem [{\citenamefont {{Babaev}}\ \emph {et~al.}(2002)\citenamefont
  {{Babaev}}, \citenamefont {{Faddeev}},\ and\ \citenamefont
  {{Niemi}}}]{babaev2001hidden}%
  \BibitemOpen
  \bibfield  {author} {\bibinfo {author} {\bibfnamefont {E.}~\bibnamefont
  {{Babaev}}}, \bibinfo {author} {\bibfnamefont {L.~D.}\ \bibnamefont
  {{Faddeev}}}, \ and\ \bibinfo {author} {\bibfnamefont {A.~J.}\ \bibnamefont
  {{Niemi}}},\ }\href {\doibase 10.1103/PhysRevB.65.100512} {\bibfield
  {journal} {\bibinfo  {journal} {\prb}\ }\textbf {\bibinfo {volume} {65}},\
  \bibinfo {eid} {100512} (\bibinfo {year} {2002})},\ \Eprint
  {http://arxiv.org/abs/cond-mat/0106152} {cond-mat/0106152} \BibitemShut
  {NoStop}%
\bibitem [{\citenamefont {Babaev}(2009)}]{babaev2009non}%
  \BibitemOpen
  \bibfield  {author} {\bibinfo {author} {\bibfnamefont {E.}~\bibnamefont
  {Babaev}},\ }\href@noop {} {\bibfield  {journal} {\bibinfo  {journal}
  {Physical Review B}\ }\textbf {\bibinfo {volume} {79}},\ \bibinfo {pages}
  {104506} (\bibinfo {year} {2009})}\BibitemShut {NoStop}%
\bibitem [{\citenamefont {Kuklov}\ and\ \citenamefont
  {Svistunov}(2003)}]{PhysRevLett.90.100401}%
  \BibitemOpen
  \bibfield  {author} {\bibinfo {author} {\bibfnamefont {A.~B.}\ \bibnamefont
  {Kuklov}}\ and\ \bibinfo {author} {\bibfnamefont {B.~V.}\ \bibnamefont
  {Svistunov}},\ }\href {\doibase 10.1103/PhysRevLett.90.100401} {\bibfield
  {journal} {\bibinfo  {journal} {Phys. Rev. Lett.}\ }\textbf {\bibinfo
  {volume} {90}},\ \bibinfo {pages} {100401} (\bibinfo {year}
  {2003})}\BibitemShut {NoStop}%
\bibitem [{\citenamefont {Kuklov}\ \emph
  {et~al.}(2004{\natexlab{a}})\citenamefont {Kuklov}, \citenamefont
  {Prokof'ev},\ and\ \citenamefont {Svistunov}}]{PhysRevLett.92.050402}%
  \BibitemOpen
  \bibfield  {author} {\bibinfo {author} {\bibfnamefont {A.}~\bibnamefont
  {Kuklov}}, \bibinfo {author} {\bibfnamefont {N.}~\bibnamefont {Prokof'ev}}, \
  and\ \bibinfo {author} {\bibfnamefont {B.}~\bibnamefont {Svistunov}},\ }\href
  {\doibase 10.1103/PhysRevLett.92.050402} {\bibfield  {journal} {\bibinfo
  {journal} {Phys. Rev. Lett.}\ }\textbf {\bibinfo {volume} {92}},\ \bibinfo
  {pages} {050402} (\bibinfo {year} {2004}{\natexlab{a}})}\BibitemShut
  {NoStop}%
\bibitem [{\citenamefont {Kuklov}\ \emph
  {et~al.}(2004{\natexlab{b}})\citenamefont {Kuklov}, \citenamefont
  {Prokof'ev},\ and\ \citenamefont {Svistunov}}]{PhysRevLett.92.030403}%
  \BibitemOpen
  \bibfield  {author} {\bibinfo {author} {\bibfnamefont {A.}~\bibnamefont
  {Kuklov}}, \bibinfo {author} {\bibfnamefont {N.}~\bibnamefont {Prokof'ev}}, \
  and\ \bibinfo {author} {\bibfnamefont {B.}~\bibnamefont {Svistunov}},\ }\href
  {\doibase 10.1103/PhysRevLett.92.030403} {\bibfield  {journal} {\bibinfo
  {journal} {Phys. Rev. Lett.}\ }\textbf {\bibinfo {volume} {92}},\ \bibinfo
  {pages} {030403} (\bibinfo {year} {2004}{\natexlab{b}})}\BibitemShut
  {NoStop}%
\end{thebibliography}%
\end{document}